\documentclass[twocolumn,preprintnumbers,amsmath,amssymb,superscriptaddress]{revtex4}

\usepackage{graphicx}
\usepackage{dcolumn}
\usepackage{bm}

\begin{document}

\preprint{}

\title{Pre- and post-selected measurements with coupling-strength-dependent modulation}
\author{Zhaoxue Li}
\author{Jiangdong Qiu}
\author{Xiaodong Qiu}
\author{Linguo Xie}
\author{Lan Luo}
\author{Xiong Liu}
\author{Yu He}
\author{Qi Wang}
\affiliation{Key Laboratory of High Energy Density Physics and Technology of Ministry of Education, Sichuan University, Chengdu 610064, China}

\author{Zhiyou Zhang\footnote{zhangzhiyou@scu.edu.cn}}
\author{JingLei Du}
\affiliation{Key Laboratory of High Energy Density Physics and Technology of Ministry of Education, Sichuan University, Chengdu 610064, China}
\affiliation{College of Physical Science and Technology, Sichuan University, Chengdu 610064, China}

\date{\today}

\begin{abstract}
Pre- and post-selected (PPS) measurement, especially the weak PPS measurement, is a useful protocol for amplifying small physical parameters. However, it is difficult to retain both the attainable highest measurement sensitivity and precision with the increase of the parameter to be measured. Here, a modulated PPS measurement scheme based on coupling-strength-dependent modulation is presented with the highest sensitivity and precision retained for an arbitrary coupling strength. This idea is demonstrated by comparing the modulated PPS measurement scheme with standard PPS measurement scheme, respectively, in the cases of balanced pointer and unbalanced pointer. By using the Fisher information metric, we derive the optimal pre- and post-selected states, as well as the optimal coupling-strength-dependent modulation without any restriction on the coupling strength. We also give the specific strategy of performing the modulated PPS measurement scheme, which may promote practical application of this scheme in precision metrology. 

\end{abstract}

\maketitle

\section{Introduction}\label{SecI}
Pre- and post-selected (PPS) measurement, which was first proposed by Aharonov, Bergmann and Lebowitz (ABL), is of fundamental significance in quantum measurement~\cite{Aharonov1964}. PPS measurement is performed on a sub-ensemble of a quantum system (QS) with chosen pre- and post-selected states such that the coupling strength between QS and measurement apparatus (MA) can be measured. A typical extension of PPS measurement is the weak PPS measurement (i.e., the so-called weak measurement), which requires an extremely weak coupling strength~\cite{Aharonov1988,Zhu2011,Kofman2012}. Weak measurement has attracted much attention due to the striking amplification effect on some ultra-small physical parameters with an amplification factor of weak value~\cite{Hosten2008,Kocsis2010,Starling2010,Zhou2012,Xu2013,Salazar-Serrano2014,Maga2014,Qiu2014,Rhee2015,Qiu2016,ZhangY2016,Xie2017,Li2018}
\begin{equation}
A_{w}=\frac{\langle\psi_{f}|\hat{A}|\psi_{i}\rangle}{\langle\psi_{f}|\psi_{i}\rangle}\label{E1},
\end{equation}
where $\hat{A}$ denotes the observable of QS. $|\psi_{i}\rangle$ and $|\psi_{f}\rangle$ denotes the pre- and post-selected states of the QS, respectively. In principle, the weak value can be very large even to far exceed the eigenvalue spectra of $\hat{A}$ when $\langle\psi_{f}|\psi_{i}\rangle\rightarrow0$\cite{Aharonov1988,Aharonov2010}. Consider a physical parameter serving as the coupling strength $g$ in a Hamilton interaction $\hat{H}=-g\hat{A}\otimes\hat{q}$ with $\hat{q}$ being the pointer of MA, the MA state after a PPS measurement can be written as
\begin{eqnarray}
\langle\psi_{f}|\Psi\rangle&=&\langle\psi_{f}|\exp(-ig\hat{A})|\psi_{i}\rangle|\Phi_{i}\rangle\nonumber\\&\simeq&\langle\psi_{f}|1-ig\hat{A}\otimes\hat{q}|\psi_{i}\rangle|\Phi_{i}\rangle\nonumber\\&\simeq&\langle\psi_{f}|\exp(-igA_{w}\hat{q})|\psi_{i}\rangle|\Phi_{i}\rangle 
\label{E2},
\end{eqnarray}
where $|\Phi_{i}\rangle $ represents the initial state of MA with a standard deviation $\sigma$. Generally, the first approximation is feasible with a weak coupling satisfying $|g|\sigma\ll1$, and the second approximation requires a weaker coupling satisfying $|gA_{w}|\sigma\ll1$\cite{Kofman2012,Duck1989,Jozsa2007,Tollaksen2010}. In this fashion, this extremely weak coupling strength $g$ can always be  extracted from the pointer shift in a linear-response regime, i.e., linear weak-value amplification with high-precision.

However, such linear weak-value amplification  is valid only for measuring extremely small parameters in the case of balanced pointer ($q_{0}=0$), such as the detection in position space\cite{Maga2014,Dixon2009}. In fact, there exists many scenarios involving unbalanced pointer which requires a nonzero expected value for initial pointer ($q_{0}\neq0$), such as the detection in frequency space\cite{Salazar-Serrano2014,Brunner2010}. For the cases of balanced and unbalanced pointer, it is difficult to retain both the attainable highest measurement sensitivity and precision with the increase of the coupling strength, even within the weak coupling limit. To address this problem, Zhang. et al recently introduced a bias phase approximating to the post-selected angle, which has the potential to retain the measurement sensitivity only for extremely weak coupling strength\cite{Zhang2016}. More recently, Li. et al proposed an adaptive weak-value amplification scheme to retain the highest precision, in which the coupling strength can be relaxed to $|g|q_{0}\ll1$\cite{Li2017}. 

In this paper, we propose a modulated pre- and post-selected measurement (PPSM) scheme based on the coupling-strength-dependent modulation. With the highest sensitivity and precision, the coupling strength can be further relaxed to an arbitrary magnitude, even to the case of strong PPS measurement. We theoretically analyze the measurement sensitivity and precision for the cases of balanced and unbalanced pointer, respectively. We also give the optimal pre- and post-selected states, as well as the optimal modulation of the coupling strength that correspond to the highest precision without any approximation. Our numerical comparison demonstrates that the modulated PPSM scheme is more feasible and efficient in precisely measuring an unknown parameter.

This paper is organized as follows. In Sec. II, we consider the common Gaussian-type wave function of MA, and theoretically analyze the modulated PPSM scheme for the cases of balanced and unbalanced pointer, respectively. In Sec. III, we take the beam deflection measurement and the time delay measurement as examples of the balanced pointer and unbalanced pointer, respectively, and compare the modulated PPSM with the standard PPSM from the view of the sensitivity, precision and post-selected probability. In Sec. IV, we give concrete realization steps of measuring an unknown parameter via the modulated PPSM scheme. We summarize our conclusions in Sec. V.

\section{Coupling-strength-dependent modulation}\label{SecII}

In this section, we begin with a generalized expression of the modulated PPS measurements. Then, we derive the optimal pre- and post-selected states, as well as the optimal coupling-strength-dependent modulation that can reach the highest measurement precision, without any restriction on the coupling strength. 

Consider a two-level QS which is pre-selected at a state of superposition $|\psi_{i}\rangle=\cos(\theta_{i}/2)|0\rangle+\sin(\theta_{i}/2)e^{i\phi_{i}}|1\rangle$, with $|0\rangle$ and $|1\rangle$ representing the eigenstates of the observable $\hat{A}=|0\rangle\langle0|-|1\rangle\langle1|$. Besides, MA is prepared at $|\Phi_{i}\rangle=\int dqf(q)|q\rangle$ in which $f(q)$ represents the wave function of MA with respect to pointer $q$. Without loss of generality, suppose $f(q)$ is the common Gaussian-type wave function centered on $q_{0}$ with a standard deviation $\sigma$, which is written as $f(q)=(\pi\sigma^{2})^{-1/4}\exp[-(q-q_{0})^{2}/2\sigma^{2}]$. The MA then is coupled to the QS with a coupling strength noted as $g$. Unlike the standard coupling in the most discussions of PPS measurement\cite{Zhu2011,Jozsa2007}, here we consider an additional modulation for the coupling strength, noted $g_{M}$. Therefore, the interaction Hamiltonian between QS and MA is described as
\begin{equation}
\hat{H}'=-(g+g_{M})\hat{A}\otimes\hat{q}\label{E3}.
\end{equation}

The QS-MA joint state then is given by
\begin{equation}
|\Psi'\rangle=e^{-i\int \hat{H}'dt}|\psi_{i}\rangle|\Phi_{i}\rangle\label{E4}.
\end{equation}

To represent the measurement precision of the coupling strength $g$ (i.e., the physical parameter to be measured), we will introduce the Fisher information (FI) metric. FI is defined based on the maximum likelihood estimation strategy, the estimation variance exists a lower limit (i.e., the Cramer-Rao bound) corresponding to the maximal FI\cite{Braunstein1994}. In principle, more FI about the parameter to be measured contained in a measurement scheme means that one can reach higher measurement precision. Thus, it is worth calculating the maximal FI contained in the pure joint state $|\Psi'\rangle$ , i.e., the so-called quantum fisher information (QFI)\cite{Braunstein1994,Braunstein1996,Pang2015}
\begin{equation}
F^{Q}=4(\langle\partial_{g}\Psi'|\partial_{g}\Psi'\rangle-|\langle\Psi'|\partial_{g}\Psi'\rangle|^{2})\label{E5}.
\end{equation}
$F^{Q}$ can reach its maximum $F^{Q}_{max}=4\langle q^{2}\rangle_{0}=4q_{0}^{2}+2\sigma^{2}$ when $\cos\theta_{i}=0$, signifying the attainable highest measurement precision (see Appendix A). Obviously, the maximal QFI is independent of the coupling strength $g$ to be measured, but only depends on the initial distribution of the MA. As the coupling strength $g$ is amplified through a post-selection on the QS, consider the post-selected state written as $|\psi_{f}\rangle=\cos(\theta_{f}/2)|0\rangle+\sin(\theta_{f}/2)e^{i\phi_{f}}|1\rangle$. In this fashion, the MA state is described as
\begin{equation}
|\Phi_{f}'\rangle=\frac{\langle\psi_{f}|\Psi'\rangle}{\sqrt{p_{d}}}\label{E6},
\end{equation}
with the successful post-selected probability $p_{d}=|\langle\psi_{f}|\Psi'\rangle|^{2}$. After the post-selection, the total intensity undergoes a large reduction while the coupling strength is significantly amplified. Now, a crucial problem arises as to whether the maximal QFI $F^{Q}_{max}$ contained in $|\Psi'\rangle$ can be completely retained after post-selection? Concretely, if only retaining the successfully post-selected MA state, the QFI contained in $|\Phi_{f}'\rangle$ is given by
\begin{equation}
F_{d}=p_{d}Q_{d}\label{E7},
\end{equation}
with $Q_{d}=4(\langle\partial_{g}\Phi_{f}'|\partial_{g}\Phi_{f}'\rangle-|\langle\Phi_{f}'|\partial_{g}\Phi_{f}'\rangle|^{2})$.
$F_{d}$ has the potential to reach $F^{Q}_{max}$ provided that $\sin\theta_{i}\sin\theta_{f}=-1$. For a given $\phi=\phi_{i}-\phi_{f}$, the maximal $F_{d}$ in the cases of balanced pointer (with $q_{0}=0$) and unbalanced pointer (with $q_{0}\neq0$ and $|q_{0}|\gg\sigma$), respectively, corresponds to the coupling strength (see Appendix B)
\begin{equation}
g_{I}=-g_{M}, g_{II}=\frac{\phi}{2q_{0}}-g_{M}\label{E8}.
\end{equation}
That is to say, the highest precision (corresponding to the maximal $F_{d}$) can be imparted to an arbitrary coupling strength $g$ in terms of the expression of Eq. (8). The maximal $F_{d}$ reaches $F^{Q}_{max}$ only when $|\phi|\ll1$. With the increase of $\phi$, the maximal $F_{d}$ tends to be decreased slightly. 

In the context of reaching the highest possible precision, the pre- and post-selected states of the QS are recast as
\begin{equation}
|\psi_{i}\rangle=\frac{1}{\sqrt{2}}(|0\rangle+|1\rangle)\label{E9},
\end{equation}
and
\begin{equation}
|\psi_{f}\rangle=\frac{1}{\sqrt{2}}(|0\rangle e^{i\phi/2}-|1\rangle e^{-i\phi/2})  \label{E10},
\end{equation}
respectively, where $\phi$ designates the post-selected angle of the QS. Therefore, by performing the modulated PPS measurements with coupling-strength-dependent modulation, the coupling strength $g$ can be extracted from the shift of the expected value of the pointer in the cases of balanced and unbalanced pointer, i.e., 
\begin{eqnarray}
\Delta\langle q\rangle_{I}&=&\frac{\int dqq|\langle q|\Phi_{f}'\rangle|^{2}}{p_{d}}\nonumber\\&=&\frac{\sigma^{2}g'\exp(-\sigma^{2}g'^{2})\sin\phi}{1-\exp(-\sigma^{2}g'^{2})\cos\phi}\nonumber\\&\simeq&\sigma^{2}g'Im(A_{w})
\label{E11},
\end{eqnarray}
and
\begin{eqnarray}
\Delta\langle q\rangle_{II}&=&\frac{\int dqq|\langle q|\Phi_{f}'\rangle|^{2}}{p_{d}}-q_{0}\nonumber\\&=&\frac{\sigma^{2}g'\exp(-\sigma^{2}g'^{2})\sin(2q_{0}g'-\phi)}{1-\exp(-\sigma^{2}g'^{2})\cos(2q_{0}g'-\phi)}\nonumber\\&\simeq&\frac{2q_{0}g'-\phi}{g'}
\label{E12},
\end{eqnarray}
respectively. $g'=g+g_{M}$ represents the total coupling strength. These two approximation terms correspond to the most precise region of the coupling strength $g$ to be measured for the cases of balanced and unbalanced pointer, respectively. For the case of balanced pointer, the approximation is feasible provided that $|g'\sigma|\ll|\phi|/2$ which is equivalent to the so-called linear region\cite{Kofman2012,Wu2011}. Under the approximation, we can see that the total coupling strength $g'$ is linearly amplified by the pointer shift with the weak value $A_{w}=i\cot(\phi/2)$. For the case of the unbalanced pointer, the approximation is feasible provided that $|g'(q_{0}+\sigma)|\sim|\phi|/2$ which refers to the so-called nonlinear intermediate region\cite{Kofman2012}.

\section{Comparison of standard and modulated PPSM schemes}\label{SecIII}
To specifically reflect the restriction of the magnitude of the coupling strength on the measurement sensitivity and precision in the standard PPSM scheme, and the advantages of the modulated PPSM scheme, we compare the two schemes in measurement sensitivity, precision and the post-selected probability.

\subsection{Balanced pointer}
We take the beam deflection measurement as an example of the case of balanced pointer, in which the transverse position (noted as $x$) is considered as the degree of freedom (DOF) of the MA. Therefore, the transverse momentum (noted as $k$) conjugate to position serves as the coupling strength to be measured in an interaction Hamilton $\hat{H}=-k\hat{A}\otimes\hat{x}$. For the modulated PPS measurement, an added transverse momentum $k_{M}$ reconstructs a modulated interaction Hamilton $\hat{H}'=-(k+k_{M})\hat{A}\otimes\hat{x}$. Moreover, the QS is considered to be a two-level system with some selectable DOFs, such as the polarization of light, the which-path of a Sagnac interferometer.

\begin{figure*}[htbp!]
\centering\includegraphics[width=17cm]{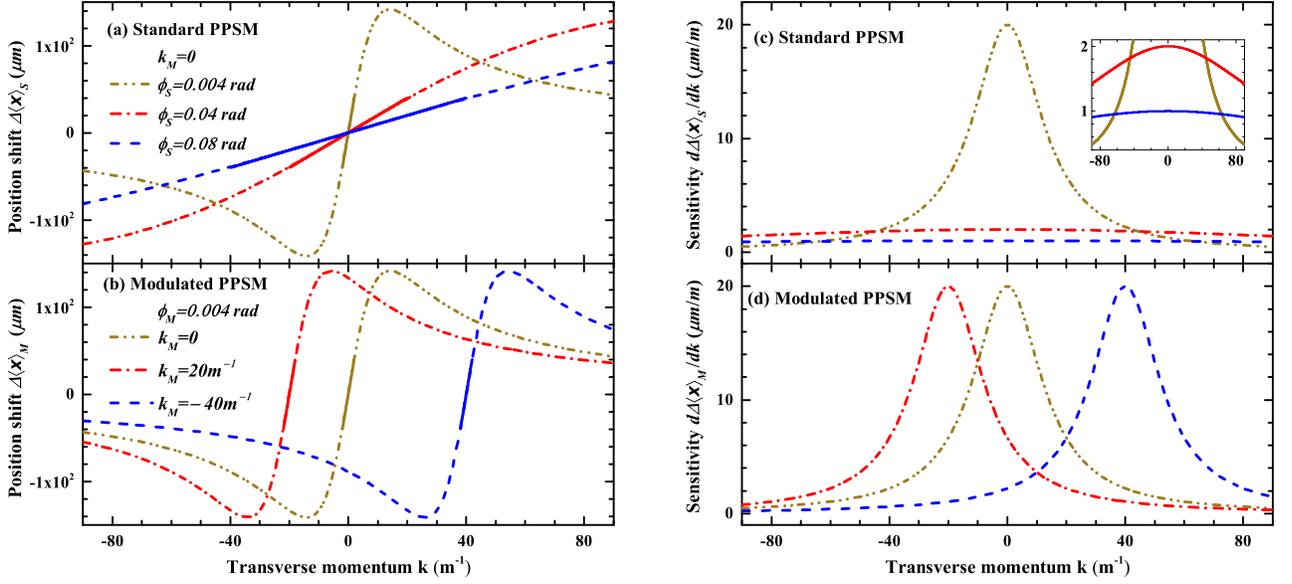}
\caption{\label{Fig1} Shift of the average transverse position and measurement sensitivity with respect to transverse momentum $k$. (a) and (c) show shift of the average transverse position with respect to transverse momentum $k$ in standard PPSM and modulated PPSM schemes, respectively. (b) and (d) show measurement sensitivity with respect to transverse momentum $k$ in standard PPSM and modulated PPSM schemes, respectively. The solid parts of the curves in (a) and (b) represent the corresponding linear regions which are respectively limited at $|k\sigma|\leq|\phi_{S}|/10$ and $|(k+k_{M})\sigma|\leq|\phi_{M}|/10$ for these two PPSM schemes.} 
\end{figure*}

In Fig. 1, we plot the shift of the average transverse position and the measurement sensitivity (characterized by the slope of the position shift curve) as a function of the transverse momentum to be measured, respectively, in standard PPSM (with $k_{M}=0$)and modulated PPSM schemes. Before the PPS measurement, a Gaussian-type wave function of the MA is centered on $x_{0}=0$ with a standard deviation assumed as $\sigma=200\mu$m. In standard PPSM scheme [see Fig. 1(a)], the position shift curve is always centered at $k=0$. For a fixed post-selected angle $\phi_{S}$, the measurable range of momentum is limited at the linear region $|k\sigma|\ll|\phi_{S}|/2$ with prominent sensitivity. With the increase of the post-selected angle, the range of the linear region can be broaden, which enables linear weak-value amplification for a lager transverse momentum. Nonetheless, the sensitivity as a whole tends to be sharply decreased with the increase of the post-selected angle, as shown in Fig. 1(c). In this regard, in the standard PPSM scheme the measurable range of momentum can be extended at the sacrifice of the sensitivity.

In modulated PPSM scheme, this issue can be solved with an appropriate modulation $k_{M}$ for transverse momentum. For a fixed post-selected angle $\phi_{M}$, the position shift curve can be arbitrarily translated and be centered at an arbitrary momentum $k=-k_{M}$, see Fig. 1(b). The modulated linear region then is limited at $|(k+k_{M})\sigma|\ll|\phi_{M}|/2$. When taking a small post-selected angle which corresponds to a narrow linear region, the high sensitivity can be retained with the increase of the momentum to be measured, see Fig. 1(d). 

To explore the influence of momentum modulation on measurement precision, we calculated the classical FI after a successful PPS measurement, which is expressed as\cite{Braunstein1996,Fisher2011}
\begin{equation}
I(k)=p_{d}\int dx P(x|k)\partial_{k}^{2}\ln P(x|k)\label{E13},
\end{equation}
where the normalized probability distribution of the momentum $k$ after a measurement on the position is expressed as $P(x|k)=|\langle x|\Phi_{f}'\rangle|^{2}$ as a function of Eq. (6), and the post-selected probability is given by $p_{d}=[1-\exp(-\sigma^{2}k^{2})\cos\phi]/2$. In the standard PPSM scheme, the profile of classical FI that is always symmetric with respect to $k=0$ is shown in Fig. 2(a). For a fixed post-selected angle $\phi_{S}$, the classical FI is mainly gathered around $k=0$ and reach the maximum $I_{max}$ at $k=0$. $I_{max}$ has the potential to reach the maximal QFI $F^{Q}_{max}=2\sigma^{2}$ at $k=0$ with an extremely small post-selected angle, meaning that the attainable highest precision may be reached only for an extremely momentum with $k\rightarrow0$. Interestingly, classical FI at the momentum $k$ except for the peak momentum region $k\rightarrow0$ tends to be increased significantly with the increase of the post-selected angle. We can see that FI at the momentum $k$ except for $k\rightarrow0$ is slightly lower than that at $k\rightarrow0$ when considering a somewhat large post-selected angle. This characteristic may be significant in measuring the continuous momentum in real time. However, FI at the peak momentum region $k\rightarrow 0$, which approximates to $I_{max}$, shows slight reduction with the increase of the post-selected angle, see the inset of Fig. 2(a). Overall the standard PPSM scheme has little ability to retain the highest measurement precision for a larger momentum. 

\begin{figure}[ht!]
\centering\includegraphics[width=8.3cm]{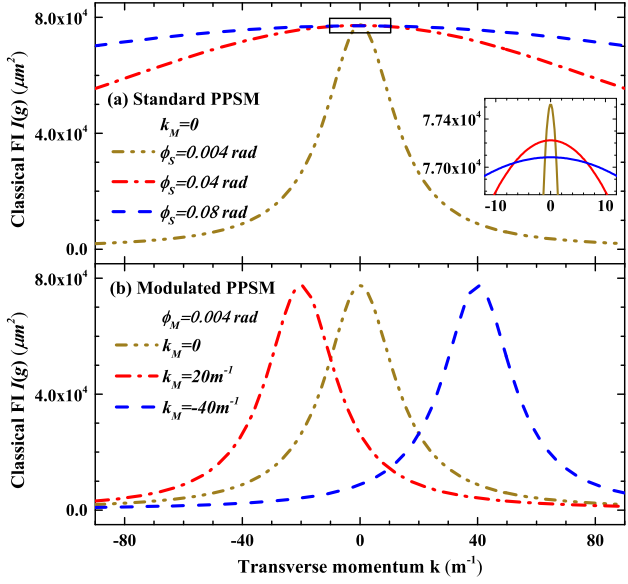}
\caption{\label{Fig2} Classical Fisher information (FI) as a function of the transverse momentum $k$ in (a) standard PPSM scheme and (b) modulated PPSM scheme, respectively.}
\end{figure}

In Fig. 2(b), we plot the corresponding classical FI curve of the modulated SPPM scheme. By performing an appropriate modulation for momentum, the FI curve can be peaked at an arbitrary momentum $k=-k_{M}$ to be measured. The peak momentum region with $k\rightarrow-k_{M}$, to some extent, corresponds to the modulated linear region with prominent sensitivity. When keeping the small post-selected angle of the standard PPSM scheme unchanged, the corresponding maximal classical FI $I_{max}$ can be always retained with the increase of the momentum to be measured. Furthermore, $I_{max}$ can reach the maximal QFI $2\sigma^{2}$ at an arbitrary momentum $k=-k_{M}$ with an extremely small post-selected angle. Therefore, using the modulated PPSM scheme has the potential to reach the highest measurement precision for an arbitrary momentum, rather than for an extremely small momentum. 

As the only restriction of post-selection on PPS measurements is the post-selected probability, a question arises as to whether the momentum modulation has an influence on post-selected probability? In Fig. 3, We compare the post-selected probability of the standard PPSM scheme and the modulated PPSM scheme. In the standard PPSM scheme [see Fig. 3(a)], the post-selected probability tends to be increased with the increase of the post-selected angle. Combined with the fact that a smaller post-selected angle can in principle enhance the measurement sensitivity and precision, in the modulated PPSM scheme we keep the small post-selected angle unchanged. As a result, the low post-selected probability [see the inset of Fig. 3(a)] is always retained for an arbitrary momentum $k=-k_{M}$, see Fig. 3(b). Therefore, in the modulated PPSM scheme the attainable highest measurement sensitivity and precision may be retained at an arbitrary momentum at the sacrifice of retaining the extremely low post-selected probability.  
\begin{figure}[ht!]
\centering\includegraphics[width=8.3cm]{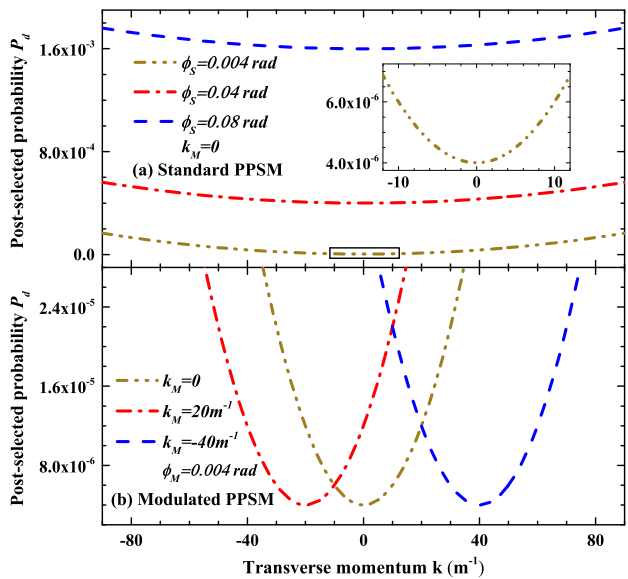}
\caption{\label{Fig3} Post-selected probability as a function of the transverse momentum $k$ in (a) standard PPSM scheme and (b) modulated PPSM scheme, respectively.}
\end{figure}

\subsection{Unbalanced pointer}
For the case of the unbalanced pointer, we take the time delay measurement for example, in which the frequency (noted as $\omega$) serves as the DOF of the MA. Suppose a Gaussian-type wave function is centered on $\omega_{0}=2400$THz with a standard deviation $\sigma=200$THz. The time delay, noted as $\tau$, serve as the coupling strength to be measured. In the modulated PPSM scheme, the  interaction Hamilton between the MA and the two-level QS is reconstructed as $\hat{H}'=-(\tau+\tau_{M})\hat{A}\otimes\hat{\omega}$ with a modulated time delay $\tau_{M}$.

\begin{figure*}[htbp!]
\centering\includegraphics[width=17.5cm]{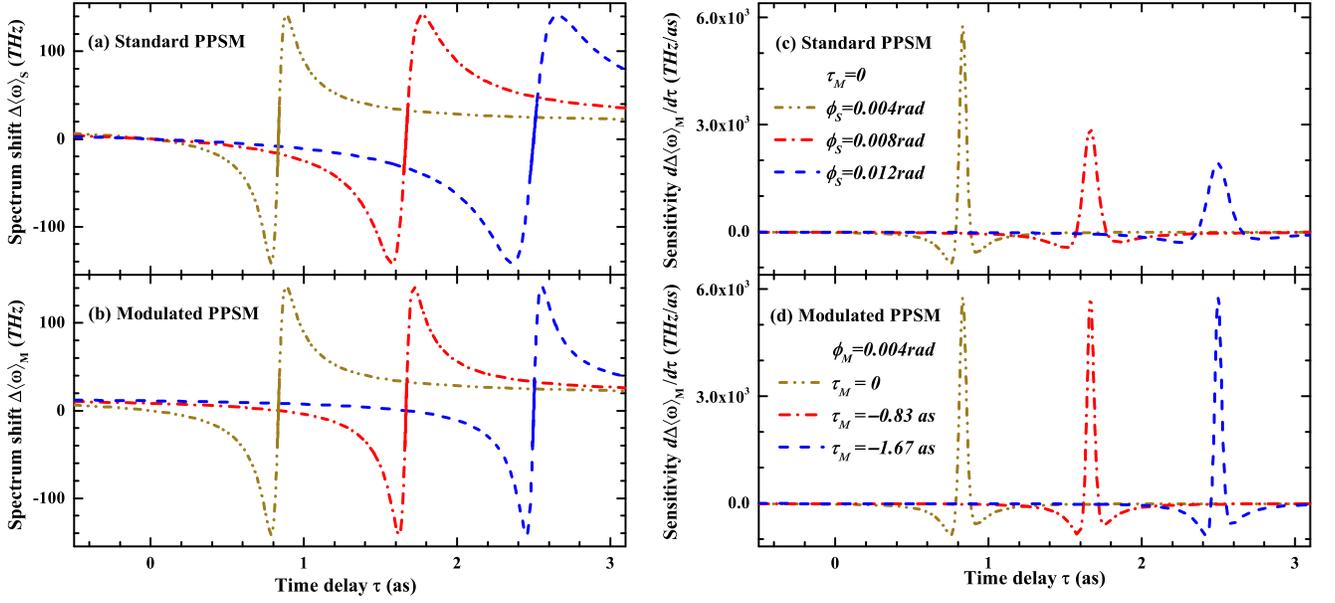}
\caption{\label{Fig4} Spectrum shift and measurement sensitivity with respect to time delay $\tau$. (a) and (b) show spectrum shift with respect to time delay $\tau$ in standard PPSM and modulated PPSM schemes, respectively. (c) and (d) show measurement sensitivity with respect to time delay $\tau$ in standard PPSM and modulated PPSM schemes, respectively. The solid parts of the curves in (a) and (b) represent the corresponding nonlinear intermediate regions which are respectively limited at $|\tau\omega_{0}-\phi_{S}/2|\leq|\tau\sigma|/10$ and $|\tau'\omega_{0}-\phi_{M}/2|\leq|\tau'\sigma|/10$ for the two PPSM schemes.} 
\end{figure*}

In Fig. 4(a) and (c), we plot the spectrum shift and the measurement sensitivity with respect to the time delay to be measured in the standard PPSM scheme, respectively. Unlike the case of balanced pointer [Fig. 1(a)] in which the pointer shift curve is always centered at the zero-value of the coupling strength, here spectrum shift curve depends on the post-selected angle $\phi_{S}$ and is centered at the time delay $\tau=\phi_{S}/2\omega_{0}$. In this case, the measurable range of the time delay is limited at the nonlinear intermediate region $|\tau\omega_{0}-\phi_{s}/2|\ll|\tau\sigma|$ which shows higher sensitivity, see Fig. 4(c). As shown in Fig. 4(a), the spectrum shift curve shows a walk-off effect along the time delay induced by the increased post-selected angle. Besides, the nonlinear intermediate region tends to be broaden with the increase of the post-selected angle, resulting in the reduction of the measurement sensitivity. Overall, standard PPSM scheme is impeded in measuring larger time delay due to sharply decreased sensitivity. 

Modulated PPSM scheme shows the superiority with the same high sensitivity retained for an arbitrary time delay, as shown in Fig. 4(d). Concretely, by adding an appropriate time-delay-dependent modulation $\tau_{M}$ and keeping the small post-selected angle $\phi_{M}$ unchanged, the spectrum shift curve can be centered on an arbitrary time delay $\tau=\phi_{M}/2\omega_{0}-\tau_{M}$, see Fig. 4(c). The walk-off effect vanishes with the fixed post-selected angle. Moreover, the measurable range of time delay is limited at the modulated nonlinear intermediate region $|\tau'\omega_{0}-\phi_{M}/2|\ll|\tau'\sigma|$ with $\tau'=\tau+\tau_{M}$.

As for the precision difference between the standard and the modulated PPSM schemes, we calculate the classical FI as follows
\begin{equation}
I(\tau)=p_{d}\int d\omega P(\omega|\tau)\partial_{\tau}^{2}\ln P(\omega|\tau)\label{E14},
\end{equation}
The normalized probability distribution with respect to time delay $\tau$ after a measurement on frequency is expressed as $P(\omega|\tau)=|\langle\omega|\Phi_{f}'\rangle|^{2}$ with the post-selected probability $p_{d}=[1-\exp(-\sigma^{2}\tau^{2})\cos(2\omega_{0}\tau-\phi)]/2$.

\begin{figure}[htbp!]
\centering\includegraphics[width=8.3cm]{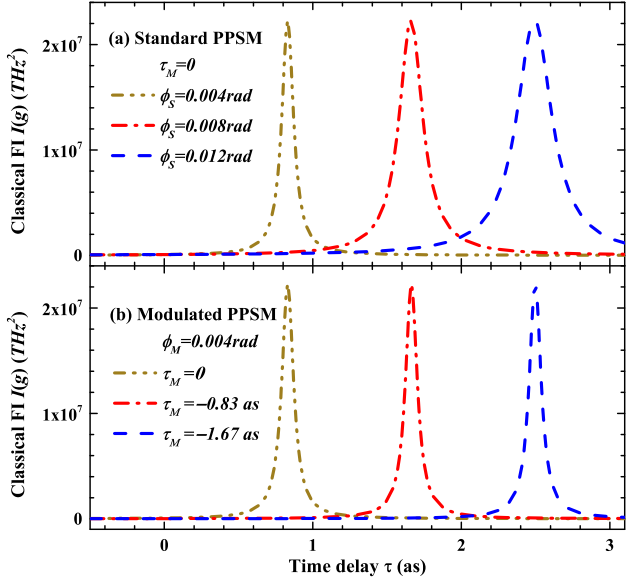}
\caption{\label{Fig5} Classical Fisher information (FI) as a function of the time delay $\tau$ in (a) standard PPSM scheme and (b) modulated PPSM scheme, respectively.}
\end{figure}

As can be seen in Fig. 5(a), the classical FI curve reaches the maximum $I_{max}$ at the peak time delay $\tau=\phi_{S}/2\omega_{0}$ for a fixed post-selected angle $\phi_{S}$. The peak time delay exactly falls within the nonlinear intermediate region. Unlike the measurement sensitivity as discussed above, it seems that $I_{max}$ (approximate to $F^{Q}_{max}=4\omega_{0}^{2}+2\sigma^{2}$) can always be retained for different time delay with the corresponding post-selected angle. In fact, $I_{max}$ tends to slightly reduced with the further increase of the post-selection angle (see Appendix B and Fig. 8), which is hardly shown in the most measurement scenarios. Furthermore, the classical FI at the time delay $\tau$ except for the peak time delay is increased significantly with the increase of the post-selected angle. That is to say, the measurable range of the time delay around $\tau=\phi_{S}/2\omega_{0}$ can be increased with an increased post-selected angle from the point of view of the measurement precision. 

For contrast, it seems that in the most scenarios the modulated PPSM scheme shows no improvement in precision, see Fig. 5(b). Combined with sensitivity curve [see Fig. 4(c) and 4(d)], we can find that the sensitivity in PPS measurements can be improved at the sacrifice of the precision corresponding to the time delay except for the peak time delay $\tau=\phi_{M}/2\omega_{0}-\tau_{M}$. However, it is more crucial for one to reach the attainable highest measurement sensitivity and precision. In this regard, modulated PPSM scheme becomes the dominate scheme to reach both the highest sensitivity and precision. 

\begin{figure}[htbp!]
\centering\includegraphics[width=8.3cm]{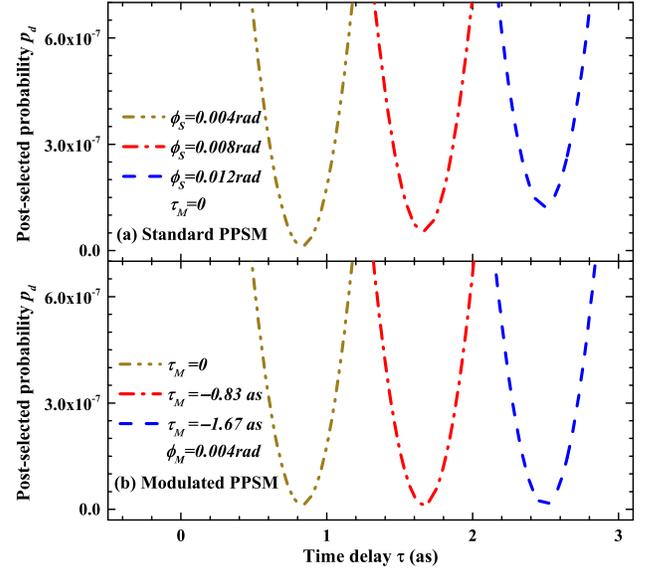}
\caption{\label{Fig6} Post-selected probability as a function of the time delay $\tau$ in (a) standard PPSM scheme and (b) modulated PPSM scheme, respectively.}
\end{figure}

In Fig. 6, we compare the post-selected probability between the standard and modulated PPSM schemes. For a fixed post-selected angle, there always exists the minimum of the post-selected probability which corresponds to the time delay $\tau$ to be measured. In the standard PPSM scheme [see Fig. 6(a)], we can see that the post-selected probability tends to be increased with the increase of the post-selected angle $\phi_{S}$. More precisely, the standard PPSM scheme shows larger post-selected probability for a larger time delay. However, in the modulated PPSM scheme the low probability is retained for an arbitrary time delay when keeping the small post-selected angle $\phi_{M}$ (corresponding to a narrow modulated intermediate region) unchanged, see Fig. 6(b).

\section{Realization of the modulated PPSM scheme}\label{SecIV}
Standard PPSM scheme shows extremely high sensitivity and precision for an extremely small coupling strength, but has little feasibility to attain the sensitivity or precision with adjustable post-selection in practice. The schematic diagram of the standard PPSM scheme is shown in Fig. 7(a). For the case of balanced pointer, the measurable range of extremely small coupling strength is limited at a narrow linear region which may miss the most precise estimation. For the case of the unbalanced pointer, the walk-off effect of the pointer shift curve hinders one from detecting the rough magnitude of the coupling strength to be measured. For contrast, modulated PPSM scheme has the ability to attain both the highest sensitivity and precision for an arbitrary coupling strength. In this case, the specific strategy of performing the modulated PPSM scheme becomes much crucial. 

\begin{figure}[ht!]
\centering\includegraphics[width=8.5cm]{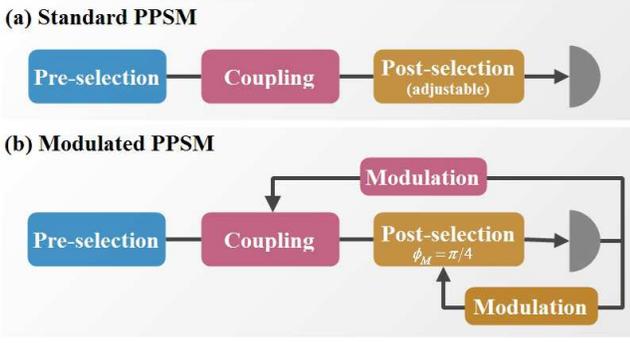}
\caption{\label{Fig7} Schematic diagram of (a) standard PPSM scheme and (b) modulated PPSM scheme. In the standard PPSM scheme, the post-selection is adjustable based on the feedback outcomes. In the modulated PPSM scheme, the post-selected angle is initially set as $\phi_{M}=\pi/4$; then, the coupling-strength-dependent modulation is introduced and the post-selection is modulated based on the feedback outcomes.}
\end{figure}

Suppose an arbitrary coupling strength (i.e., the parameter to be measured) is expressed as $g=g_{0}+\Delta g$ with $g_{0}$ and $\Delta g$ representing the expected value of measurement outcomes and the uncertainty, respectively. More precisely, the uncertainty $\Delta g$ can be regarded as a small fluctuation of measured coupling strength which is not detectable in general. Modulated PPSM scheme shows its significance in measuring $\Delta g$ with high-precision, with a schematic diagram shown in Fig. 7(b). Concretely, the modulated PPSM scheme in the case of balanced pointer is performed in terms of three steps: first, the coupling strength can be roughly estimated as $g_{0}$ with a fixed post-selected angle $\phi_{M}=\pi/4$ which is analogous to the balanced homodyne detection\cite{Qiu2017}; then, the coupling-strength-dependent modulation is determined as $g_{M}=-g_{0}$ and the post-selected angle is modulated based on $\Delta g$ ($|\Delta g\sigma|\ll|\phi_{M}|/2$), i.e., $\Delta g$ is located within the modulated linear region $|(g+g_{M})\sigma|\ll\phi_{M}|/2$; finally, the uncertainty $\Delta g$ is precisely estimated.

In the case of the unbalanced pointer, the modulated PPSM scheme is also divided into three steps: first, the post-selected angle is chosen as $\phi_{M}=\pi/4$ and the coupling-strength-dependent modulation is set as $g_{M}=\phi_{M}/8q_{0}$, the coupling strength is roughly estimated as $g_{0}$; then, the coupling-strength-dependent modulation is determined as $g_{M}=\phi_{M}/2q_{0}-q_{0}$ and the post-selected angle is modulated based on $\Delta g$ ($|\Delta gq_{0}-\phi_{M}/2|\ll|\Delta g\sigma|$) as a function of the feedback information of the rough estimation, i.e., $\Delta g$ is located within the modulated nonlinear intermediate region $|g'q_{0}-\phi_{M}/2|\ll|g'\sigma|$; finally, the uncertainty  $\Delta g$ is precisely estimated.

\section{Conclusions}

By including appropriate modulation for the coupling strength to be measured and the post-selection state, we are able to reach the attainable highest sensitivity and precision in the modulated PPSM scheme. We have given the optimal modulation corresponding to the optimal sensitivity and precision, which meets precision measurement for an arbitrary coupling strength without any approximation. 

In this work, we have compared the modulated PPSM scheme to the standard PPSM scheme, respectively, in the case of the balanced pointer and unbalanced pointer. For the case of balanced pointer, both sensitivity and precision tends to be decreased with the increase of the coupling strength to be measured in the standard PPSM scheme. Modulated PPSM scheme has the ability to retain high sensitivity and precision for an arbitrary coupling strength only at the cost of retaining low post-selected probability. For the case of the unbalanced pointer, both the highest sensitivity and precision may be reached only via the modulated PPSM scheme, which inevitably requires low post-selected probability. As for the implement of the modulated PPSM scheme in practical experiments, we have given the specific strategy of performing the modulated PPSM scheme.

\begin{acknowledgements}
We are sincerely grateful to the anonymous referee, whose comments
have led to a significant improvement of our paper. This research
was supported by Natural Science Foundation of China (Grant No. 11674234).
\end{acknowledgements}

\begin{appendix} 
\section{Quantum Fisher information of $|\Psi'\rangle
$}
Consider a two-level QS pre-selected at the state $|\psi_{i}\rangle=\cos(\theta_{i}/2)|0\rangle+\sin(\theta_{i}/2)e^{i\phi_{i}}|1\rangle$, and a MA initially prepared at $|\Phi_{i}\rangle=\int dqf(q)|q\rangle$ with the wave function $f(q)=(\pi\sigma^{2})^{-1/4}\exp[-(q-q_{0})^{2}/2\sigma^{2}]$. An additional coupling-strength-dependent modulation, noted as $g_{M}$, reconstructs the interaction Hamilton as $\hat{H}'=-(g+g_{M})\hat{A}\otimes\hat{q}$. The QS-MA joint state then is expressed as

\begin{equation}
\begin{split}
|\Psi'\rangle=&e^{-i\int\hat{H'}dt}|\psi_{i}\rangle|\Phi_{i}\rangle\\=&\int dq[\cos\frac{\theta_{i}}{2}|0\rangle e^{i(g+g_{M})q}\\&+\sin\frac{\theta_{i}}{2}|1\rangle e^{i\phi_{i}}e^{-i(g+g_{M}q)}]f(q)|q\rangle
\label{A1},
\end{split}
\end{equation}

The quantum Fisher information (QFI) carried by $|\Psi'\rangle$ is calculated as follows
\begin{equation}
F^{Q}=4(\langle\partial_{g}\Psi'|\partial_{g}\Psi'\rangle-|\langle\Psi'|\partial_{g}\Psi'\rangle|^{2})
\label{A2}.
\end{equation}
where
\begin{eqnarray}
\langle\partial_{g}\Psi'|\partial_{g}\Psi'\rangle=\int dq q^{2}f^{2}(q)=q_{0}^{2}+\frac{1}{2}\sigma^{2}
\label{A3},
\end{eqnarray}

\begin{eqnarray}
\langle\Psi'|\partial_{g}\Psi'\rangle=i\cos\theta_{i}\int dq qf^{2}(q)=i\cos\theta_{i}q_{0}
\label{A4}.
\end{eqnarray}
Substituting Eqns. (A3-A4) into Eq. (A2), we have $F^{Q}=4q_{0}^{2}+2\sigma^{2}-4\cos^{2}(\theta_{i})q_{0}^{2}$. $F^{Q}$ reaches the maximum $F^{Q}_{max}=4q_{0}^{2}+2\sigma^{2}$ when $\cos\theta_{i}=0$.

\section{Quantum Fisher information of $|\Phi'_{f}\rangle
$}
Suppose the QS is post-selected at $|\psi_{f}\rangle=\cos(\theta_{f}/2)|0\rangle+\sin(\theta_{f}/2)e^{i\phi_{f}}|1\rangle$, the normalized MA state then is written as
\begin{eqnarray}
|\Phi'_{f}\rangle&=&\frac{\langle\psi_{f}|\Psi'\rangle}{\sqrt{p_{d}}}
\nonumber\\&=&\frac{1}{\sqrt{p_{d}}}\int dq[\cos\frac{\theta_{i}}{2}\cos\frac{\theta_{f}}{2}e^{i(g+g_{M})q}\nonumber\\&&+\sin\frac{\theta_{i}}{2}\sin\frac{\theta_{f}}{2}e^{i\phi}e^{-i(g+g_{M})q}] f(q)|q\rangle
\label{B1}.
\end{eqnarray}
Under the condition of reaching the maximal QFI of the state $|\Psi'\rangle$ (with $\cos\theta_{i}=0$), the probability of successful post-selection is given by
\begin{equation}
p_{d}=\frac{1\pm\sin\theta_{f}\int dq\cos[2(g+g_{M})q]f^{2}(q)}{2}
\label{B2},
\end{equation}
where $\pm$ corresponds to that $\sin\theta_{i}=\pm1$.
Hence, the normalized MA state is rewritten as
\begin{equation}
|\Phi'_{f}\rangle=\frac{\int dq[\cos\frac{\theta_{f}}{2}e^{i(g+g_{M})q}\pm\sin\frac{\theta_{f}}{2}e^{i\phi}e^{i(g+g_{M})q}]f(q)|q\rangle}{\sqrt{2}\xi(g)}
\label{B3},
\end{equation}
where we define $\xi(g)=\sqrt{p_{d}}$ for later analysis. Then, we have

\begin{equation}
\begin{split}
|\partial_{g}\Phi'_{f}\rangle=&\frac{1}{\sqrt{2}\xi^{2}(q)}\int dq\{\cos\frac{\theta_{f}}{2}e^{ig'q}[iq\xi(g)-\xi'(g)]\\&\pm\sin\frac{\theta_{f}}{2}e^{i\phi}e^{-ig'q}[-iq\xi(g)-\xi'(g)]\}f(q)|q\rangle
\label{B4},
\end{split}
\end{equation}

\begin{equation}
\begin{split}
\langle\partial_{g}\Phi'_{f}|\partial_{g}\Phi'_{f}\rangle=&\frac{1}{2\xi^{4}(g)}\int dq\{q^{2}\xi^{2}(g)+\xi'^{2}(g)\\&\pm[-q^{2}\xi^{2}(g)+\xi'^{2}(g)]\sin\theta_{f}\cos(2g'q-\phi)\\&\pm q\xi(g)\xi'(g)\sin\theta_{f}\sin(2g'q-\phi)\}f^{2}(q)
\label{B5},
\end{split}
\end{equation}

\begin{equation}
\begin{split}
\langle\Phi'_{f}|\partial_{g}\Phi'_{f}\rangle=&\frac{1}{2\xi^{3}(g)}\int dq\{iq\xi(g)\cos\theta_{f}-\xi'(g)\mp\sin\theta_{f}[q\\&\xi(g)\sin(2g'q-\phi)+\xi'(g)\cos(2g'q-\phi)]\}f^{2}(q)
\label{B6},
\end{split}
\end{equation}
where $g'=g+g_{M}$ denotes the total coupling strength.

For the case of balanced pointer (with $q_{0}=0$), the QFI of the state $|\Phi'_{f}\rangle$ is calculated as

\begin{equation}
\begin{split}
F_{d}=&4p_{d}(\langle\partial_{g}\Phi'_{f}|\partial_{g}\Phi'_{f}\rangle-|\langle\Phi'_{f}|\partial_{g}\Phi'_{f}\rangle|^{2})\\\simeq&\sigma^{2}(1+\cos\phi)+2(1-\cos\phi)\xi_{0}'^{2}(g)/\xi_{0}^{2}(g)\\&-(1-\cos\phi)^{2}\xi_{0}'^{2}(g)/\xi_{0}^{4}(g)\\\simeq&2\sigma^{2}
\label{B7},
\end{split}
\end{equation}
where the first approximation is feasible with $|(g+g_{M})\sigma|\ll1$, $\xi_{0}(g)$ and $\xi_{0}'(g)$ denote the corresponding $\xi(g)$ and $\xi'(g)$ under this approximation, respectively. The second approximation is feasible with $|\phi|\ll1$. Under these limits, the maximal QFI $F^{Q}_{max}=2\sigma^{2}$ of $|\Psi'\rangle$ can be completely retained after post-selection. Note that these approximations are derived provided that $\sin\theta_{i}\sin\theta_{f}=-1$.

For the case of unbalanced pointer (with $q_{0}\neq0$ and $|q_{0}\gg\sigma|$), the QFI of the state $\Phi'_{f}$ is calculated as
\begin{equation}
\begin{split}
F_{d}=&4p_{d}(\langle\partial_{g}\Phi'_{f}|\partial_{g}\Phi'_{f}\rangle-|\langle\Phi'_{f}|\partial_{g}\Phi'_{f}\rangle|^{2})\\\simeq&(2q_{0}^{2}+\sigma^{2})(1+\eta)-2\eta g'^{2}\sigma^{4}+2(1-\eta)\zeta'^{2}(g)/\zeta^{2}(g)\\&-2\eta g'\sigma^{2}\zeta'(g)/\zeta(g)-(1-\eta g'\sigma^{2}-\eta)^{2}\zeta'^{2}(g)/\zeta^{4}(g)\\\simeq&4q_{0}^{2}+2\sigma^{2}
\label{B8},
\end{split}
\end{equation}
\begin{figure}[ht!]
\centering\includegraphics[width=8.5cm]{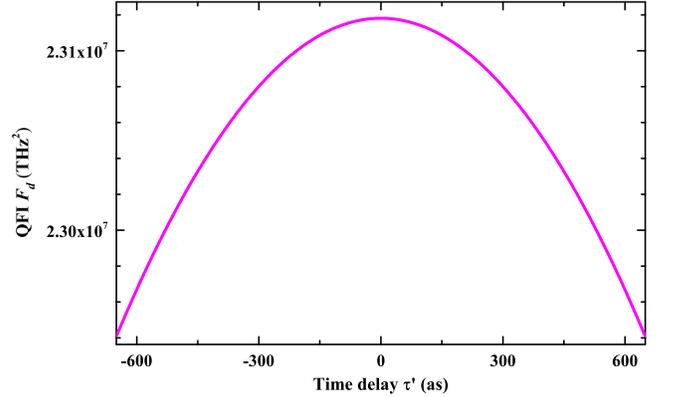}
\caption{\label{Fig8} Quantum Fisher information (QFI) of the state $|\Phi'_{f}\rangle$ as a function of the total time delay $\tau'=\tau+\tau_{M}$ under the condition that $\phi_{M}=2(\tau+\tau_{M})\omega_{0}$.}
\end{figure}
where $\eta=\exp(-g'^{2}\sigma^{2})$. The first approximation is feasible with $\phi=2(g+g_{M})q_{0}$, $\zeta(g)$ and $\zeta'(g)$ denote the corresponding $\xi(g)$ and $\xi'(g)$ under this approximation, respectively. The second approximation is feasible with $|(g+g_{M})\sigma|\ll1$. Obviously, the maximal QFI $F^{Q}_{max}=4q_{0}^{2}+2\sigma^{2}$ of the state $|\Psi'\rangle$ can be completely retained under these approximations. Fig. 8 shows the influence of the total coupling strength $g'=g+g_{M}$ on the QFI of the state $|\Phi'_{f}\rangle$ under the condition that $\phi=2(g+g_{M})q_{0}$, which is exampled by the time delay measurements (with $\tau'$ serving as $g'$) discussed in the text. We can see that $F_{d}$ undergoes only a slight reduction for a large range of the time delay $\tau'$. In the modulated PPSM scheme, by introducing an appropriate modulation $\tau_{M}$ to meet $|\tau'\sigma|\ll1$, the maximal FI $F^{Q}_{max}$ can always be retained for an arbitrary time delay $\tau$ to be measured. In analogy to the standard PPSM scheme ($\tau_{M}=0$), $F_{d}$ tends to be slightly reduced with the increase of the time delay $\tau$ to be measured. For the case of the unbalanced pointer, this reduction of QFI is hardly shown in the most measurement scenarios.

\end{appendix}

\end{document}